\documentclass[12pt]{article}
\usepackage[margin=1.25in]{geometry}

\usepackage{xcolor} 
\definecolor{kellygreen}{RGB}{50, 143, 75}
\usepackage[colorlinks=true, allcolors=kellygreen]{hyperref} 
\usepackage{graphicx} 
\usepackage{ulem} 
\usepackage{bm} 
\usepackage{floatrow} 
\usepackage[font=footnotesize, labelfont=bf]{caption} 
\usepackage{nicefrac}       
\usepackage{placeins} 
\usepackage[inline]{enumitem} 
\usepackage{newtxtext,newtxmath} 
\usepackage{natbib} 
\setcitestyle{aysep={}} 
\usepackage{xurl} 

\usepackage{authblk}

\usepackage{amsfonts,amsmath}
\numberwithin{equation}{section} 

\newcommand{\R}{\mathcal{R}}
\newcommand{\pcrit}{p_{\rm crit}}
\newcommand{\scrit}{s_{\rm crit}}

\newcommand{\eg}{\emph{e.g.,} }
\newcommand{\ie}{\emph{i.e.,} }

\newcommand{\includepdfw}[2]{\includegraphics[width=#1\textwidth]{#2.pdf}}

\newcommand{\elab}[1]{\label{eqn:#1}}
\newcommand{\flab}[1]{\label{fig:#1}}
\newcommand{\slab}[1]{\label{sec:#1}}

\newcommand{\eref}[1]{{\color{black}Equation~\ref{eqn:#1}}}
\newcommand{\fref}[1]{{\color{black}Figure~\ref{fig:#1}}}
\newcommand{\sref}[1]{{\color{black}Section~\ref{sec:#1}}}
\newcommand{\frefp}[1]{{\color{black}(Figure~\ref{fig:#1})}}

\title{Modeling the interplay between seasonal flu outcomes and individual vaccination decisions}

\author[1]{Irena Papst*}
\author[2]{Kevin P. O'Keeffe}
\author[3]{Steven H. Strogatz}
\affil[1]{Center for Applied Mathematics, Cornell University, Ithaca, NY 14853}
\affil[2]{Senseable City Laboratory, Massachusetts Institute of Technology, Cambridge, MA 02139}
\affil[3]{Department of Mathematics, Cornell University, Ithaca, NY 14853}

\begin{document}
\maketitle

\vspace{-2em}

\begin{center}
    \footnotesize *Corresponding author: \href{mailto:ip98@cornell.edu}{ip98@cornell.edu}
\end{center}

\vspace{1em}

\begin{abstract}
Seasonal influenza presents an ongoing challenge to public health. The rapid evolution of the flu virus necessitates annual vaccination campaigns, but the decision to get vaccinated or not in a given year is largely voluntary, at least in the United States, and many people decide against it. In early attempts to model these yearly flu vaccine decisions, it was often assumed that individuals behave rationally, and do so with perfect information---assumptions that allowed the techniques of classical economics and game theory to be applied. However, the usual assumptions are contradicted by the emerging empirical evidence about human decision-making behavior in this context. We develop a simple model of coupled disease spread and vaccination dynamics that instead incorporates experimental observations from social psychology to model annual vaccine decision-making more realistically. We investigate population-level effects of these new decision-making assumptions, with the goal of understanding whether the population can self-organize into a state of herd immunity, and if so, under what conditions. Our model agrees with established results while also revealing more subtle population-level behavior, including biennial oscillations about the herd immunity threshold.
\end{abstract}

\noindent\hspace{2.2em} \textbf{Keywords:} seasonal influenza, vaccination, decision-making, social

\noindent\hspace{2.2em} psychology, SIR model

\newpage

\setlength{\parskip}{1em}

\section{Introduction}

Annual influenza epidemics are a significant public health challenge, with up to 650,000 individuals dying from respiratory diseases associated with the flu each year \citep{WHOnews}. In the United States alone, the total economic burden of seasonal influenza, including direct medical costs and lost earnings due to illness or death, has been estimated as \$26.8 billion annually \citep{Molin+07}.

One of the main challenges of controlling seasonal influenza spread is that the viruses evolve quickly (on the same time scale as the annual epidemics), with multiple strains circulating concurrently. A key adaptation  mechanism, antigenic drift, gives rise to new influenza strains by randomly changing segments of viral surface proteins. Given that a host's immune system uses these surface proteins to identify the virus so that it may be neutralized \citep{TaubeKash10}, antigenic drift thus acts as an evolutionary countermeasure. It  helps the flu evade the immune system and thereby promotes its spread through the host population.

The rapid evolution of the flu results in the constant threat of a pandemic, and it also makes it challenging to develop effective, long-lasting vaccines. The seasonal flu vaccine is updated every year to protect against the strains that seem to pose the largest upcoming threat. Seasonal influenza vaccination is largely voluntary in the United States, so individuals must decide whether or not to vaccinate each year.

For many people, this decision is not easy. It involves many quantities that are effectively impossible for an individual to estimate accurately, such as their likelihood of being vaccinated successfully, the probability of an adverse reaction to the vaccine, as well as their increased risk of catching the flu by foregoing the vaccine. Yet although the decision may be difficult at an individual level, at a societal level the benefits of vaccinating can be immense; if a critical mass of individuals choose to immunize themselves, ``herd immunity'' can be achieved. In this desirable state, the density of susceptible individuals is so low that an infection chain cannot be sustained and so an epidemic cannot occur \citep{Fine93}.

Somewhat paradoxically, the possibility of achieving herd immunity makes an individual's decision to vaccinate or not even more complex, at least when viewed through the lens of classical game theory (wherein agents are assumed to be purely self-interested, and to behave rationally with perfect information about the situation at hand). The issue is that as vaccination coverage increases, individuals are increasingly incentivized \emph{not} to vaccinate. Each person would do better relying on others to bear any burden associated with vaccination, while everyone reaps the benefits of widespread immunity. Such ``free-riding'' logic makes it impossible to ever actually achieve herd immunity with rational hosts.

The free-riding problem is common in models where agents rationally weigh a delayed collective group benefit against immediate individual costs. Individuals are assumed to make decisions by selecting the strategy that maximizes an objective, individual payoff function, as prescribed by classical economics \citep{Hardi13}. Some early models of voluntary vaccination decisions involve such assumptions and inevitably yield agents that utilize free-riding logic, which precludes herd immunity \citep{GeoffPhili97,BauchEarn2004}.

However, a more recent empirical study suggests that free-riding logic is uncommon when individuals specifically consider whether or not to get the seasonal influenza vaccine \citep{Parke+13}. In fact, this study finds that the majority of individuals surveyed do not account for the vaccination decisions of others when making their own decision. Nevertheless, despite increasing evidence that assumptions from classical economics may not appropriately capture human behavior in the context of infectious disease spread, some behavior-disease models continue to be built on such foundations \citep{Verel+16}.

Other studies have challenged these assumptions by replacing them with those from behavioral economics, which leverages social psychology in its models of human decision-making. \citet{Voins+15} develop a behavior-disease model that incorporates cognitive biases and differing vaccine opinions among individuals to study vaccination coverage over time. \citet{OrabyBauch15} study pediatric acceptance of vaccines by incorporating prospect theory into their disease model.

In this paper, we consider a simplified model for the interplay between annual vaccination decisions and seasonal influenza spread, in which individual voluntary vaccination decisions are informed by observed social psychology in this context. Unlike previous work, we model \emph{repeated} vaccination decisions to reflect the annual vaccination decision necessitated by the rapid evolution of influenza viruses. We investigate population-level effects of these new decision-making assumptions, with the goal of understanding whether the population can self-organize into a state of herd immunity, and if so, under what conditions. Despite our model's idealized nature, we find that its results align with those utilizing assumptions based on classical economics, although our model also predicts more nuanced population-level behaviors, such as oscillations in and out of herd immunity on a biennial basis.

\section{Conceptual Model}
\slab{concep-model}

Decision theory and social psychology suggest that, in general, individuals tend to use heuristics, or rules of thumb, rather than a  ``rational'' cost-benefit analysis in complex decision-making \citep{TversKahne74}. Moreover, decision-making tends to obey the law of inertia: choices generally remain unchanged but are sensitive to both small nudges and unfavorable resulting outcomes \citep{ThaleSunst09}. Our model is based on both of these ideas.

For simplicity, assume that each year, an individual chooses whether or not to receive the seasonal influenza vaccine based solely on evaluating their most recent outcome with both the vaccine and the disease. The vaccine carries a cost, or risk, of adverse reaction, which can be interpreted as a cost to an individual's health due to vaccine side effects (morbidity), a direct economic cost from paying for the vaccine, and/or an indirect economic cost such as taking unpaid leave from work to get vaccinated. In what follows, we will interpret vaccine cost as morbidity, but the modeling framework is flexible enough to accommodate other interpretations. There is also some probability that vaccination successfully confers immunity upon the recipient.

Before we present the model in detail in \sref{math-model}, let us first describe it intuitively. In any given year, individuals make the decision of whether or not to vaccinate, then follow through with their choice, and then the flu season occurs. The epidemic resolves itself, each individual assesses their personal outcome from the past year, and then decides whether or not to get vaccinated prior to next year's flu season. The decision rule is a simple heuristic: if a person ``won'' last year (did not get sick and did not have an adverse reaction to the vaccine), they stick to their vaccination choice and make the same decision the following year. If they ``lost'' (got sick or had a bad reaction to the vaccine), they are nudged to change their behavior (switch from vaccinating to not, or vice versa) in hopes of eliciting a better during the next flu season.

\begin{figure}
\floatbox[{\capbeside\thisfloatsetup{capbesideposition={right,center},capbesidewidth=0.45\textwidth}}]{figure}[\FBwidth]
{\caption{\textbf{All possible decisions and outcomes for an individual in year $\bm{n}$, leading to different decisions for year $\bm{n}+\bm{1}$.} Boxes with solid borders denote decisions and possible intermediate repercussions in year $n$. Boxes with dashed and dotted borders denote final repercussions that determine the decision in year $n+1$: vaccinate (box with dashed border) or do not vaccinate (box with dotted border). In the model, $p_n$ is the proportion of vaccinators in year $n$, $r$ is the probability that the vaccine induces a cost, $s$ is the probability that the vaccine succeeds, $\phi$ is the final size of the epidemic (normalized as a fraction of the whole population), and $f(\phi) = \phi(sp_n)/(1-sp_n)$ is the fraction of the susceptible population that was infected in the  epidemic occurring in year $n$. A ``bad reaction'' results when one incurs a cost from the vaccine (\eg vaccine side effects on health, or economic burden). See \sref{math-model} for model details.}\flab{decision-flowchart}}
{\includepdfw{0.55}{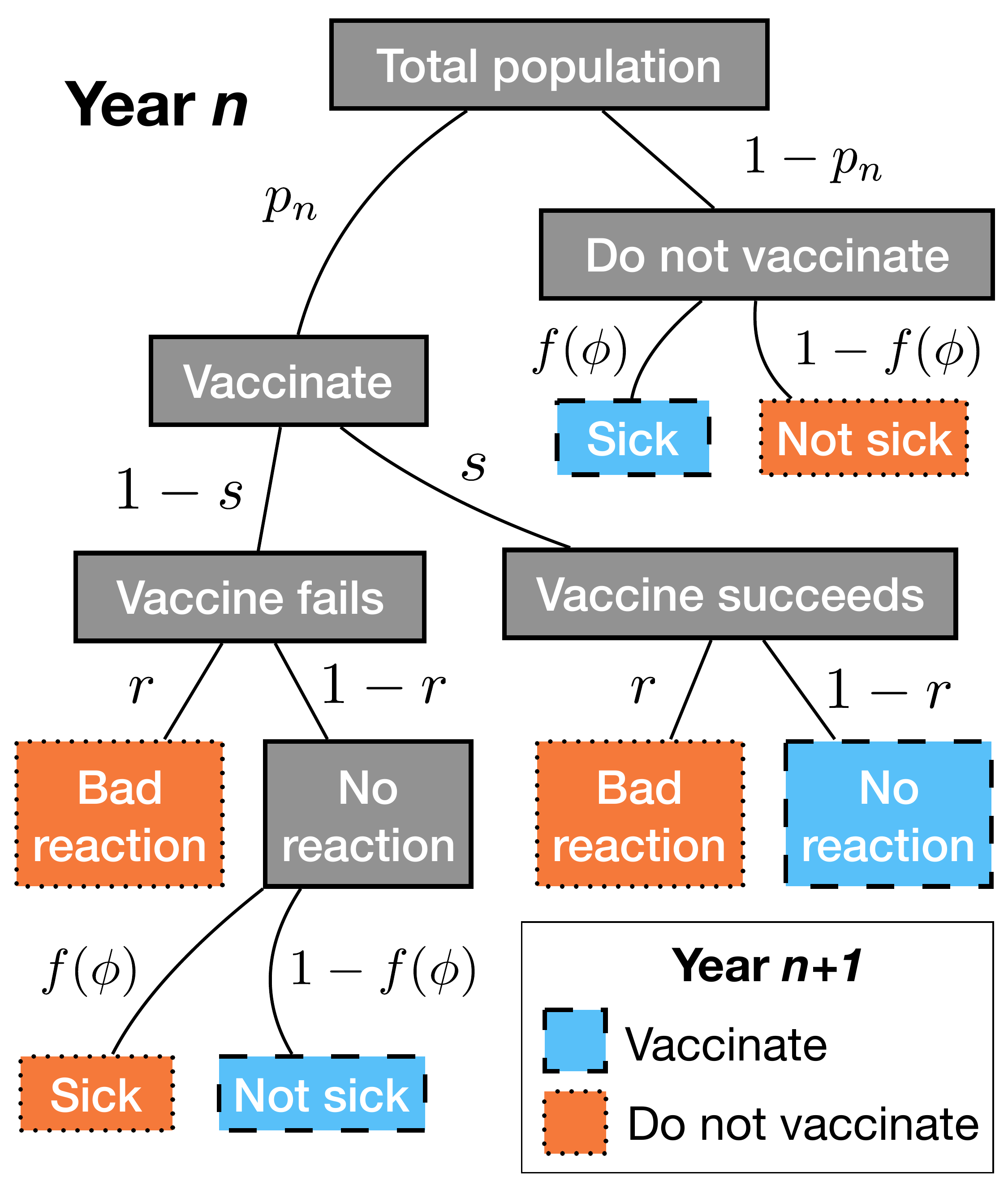}}
\end{figure}

\fref{decision-flowchart} shows a schematic of the model. As an example of how to flow through the chart, let us first consider an individual who has decided to vaccinate in a given year (a decision made by a proportion $p_n$ of the population in year $n$). This decision corresponds to the top left fork of the tree in \fref{decision-flowchart}. Next, suppose the vaccine succeeds in conferring immunity (which occurs with probability $s$), and the vaccine does not elicit an adverse reaction (which occurs with probability $1-r$). This favorable outcome does not push the individual away from their default (winning) strategy of vaccinating, so they decide to vaccinate again the following year; the vaccine seems to have succeeded in protecting them from the flu. (We are assuming here that only the failure of the vaccine can be observed by the individual, and only if they happen to get ill that year. Otherwise, the success of the vaccine is presumed, since there is no evidence to the contrary.)

On the other hand, an individual may choose \emph{not} to vaccinate in a given year (a decision made by a proportion $1-p_n$ of the population, shown by following the top right fork in \fref{decision-flowchart}). If such an individual then happens to contract the flu (which occurs with a probability that we will calculate below), then since this non-vaccinating individual's choice was a ``losing'' strategy, they decide to vaccinate the following year. The various other paths through the tree can be understood similarly.

\newpage

In the next section, we write down the governing equations for our model. We begin by recalling some standard results for a simple epidemiological model and then couple that model to a social psychological model for individual vaccination decisions.

\section{Mathematical Model}
\slab{math-model}

\subsection{Epidemic model}

We choose the susceptible-infected-removed (SIR) model of infectious disease dynamics \citep{KermaMcKen27}, both because it is a well-established epidemiological model and because there exists an analytical expression for the final size of epidemics predicted by the model. The SIR model is adequate for modeling each annual influenza outbreak individually, though we note that a more realistic flu model could be substituted into our vaccination coverage model, provided that the final size could at least be calculated numerically.

Suppose $S(t)$ and $I(t)$ represent proportions of a population that are susceptible and infected, respectively, at a time $t$. Infected individuals are assumed to be immediately infectious; there is no latency period in this  model. There may also be individuals that are removed from the infection process as they have already recovered from the illness (denoted by proportion $R(t)$), but since we assume this disease propagates in a closed population, we have $R(t) = 1 - S(t) - I(t)$, which means we do not need to track the removed individuals explicitly.

The SIR model is defined by a set of two coupled, nonlinear, ordinary differential equations:
\begin{subequations}
  \begin{align}
    \frac{dS}{dt} &= -\beta SI,  \\
    \frac{dI}{dt} &= \beta SI - \gamma I,
  \end{align}
\end{subequations}

\noindent where $\beta$ is the disease transmission rate and $\gamma$ is the disease recovery rate. The derived quantity $\R_0 = \beta/\gamma$ is the basic reproduction number of the disease; it gives the average number of secondary cases generated by an infectious individual in a fully susceptible population over the course of their illness.

In our model, we assume the initial conditions

\begin{equation}
S(0) = 1-sp, \quad 0< I(0) << 1, \quad R(0) = sp
\end{equation}

\noindent to incorporate a vaccine uptake level at proportion $p$ with success probability $s$.

\subsection{Final size of the epidemic}

For this model, one can derive an implicit equation for the final size of the epidemic $\phi(x)$, where $x$ is the proportion initially immune to the disease \citep{MaEarn06}:

\begin{equation}
\phi(x) =(1-x)(1-e^{-\R_0\phi(x)}).
\elab{phi-impl}
\end{equation}

The solution to \eref{phi-impl} can be written in terms of the principal branch of the product log function (\ie the Lambert W-function), denoted by $W[ \cdot ]$:

\begin{equation}
\phi(x) = 1-x + \frac{1}{\R_0}W\left[-\R_0(1-x)e^{-\R_0(1-x)}\right]
\elab{phi-expl}
\end{equation}

This expression for the final size of the yearly influenza epidemic is used later in \eref{model-map} to complete the model.

\subsection{Critical vaccination threshold for herd immunity}

An epidemic cannot be sustained if the average number of secondary cases provoked by an infected individual in the population is below one (since this infected individual cannot even replace themselves in the infection chain, let alone generate further infections). In other words, for the population to achieve herd immunity, the \emph{effective} reproduction number (the basic reproduction number times the proportion currently susceptible), $\R_{\rm eff} = \R_0(1-sp)$, must be driven below 1. Thus the critical vaccination threshold, $\pcrit$, satisfies the equation $\R_0(1-s\pcrit) = 1$, and so

\begin{equation}
    \pcrit = \frac{1}{s} \left( 1-\frac{1}{\R_0} \right).
\end{equation}

\subsection{Estimating \texorpdfstring{{$\R_0$}}\xspace \hspace{0.1em} for seasonal influenza}

Estimates of $\R_0$ for vary depending on year, location, and influenza subtype since the basic reproduction number depends not only on the immunological properties of the virus, but also on the social behavior of the host population. A systematic review by \citet{Bigge+14} catalogues many estimates of both the basic and effective reproduction numbers of pandemic, zoonotic, and seasonal influenza.

The most relevant estimates for our study are those for the 1976-1981 outbreak of H1N1/H3N2/B in the USA. Two studies were performed to estimate the basic reproduction number in this outbreak, and they both use serologically confirmed infections for their data, which make these estimates particularly reliable\footnote{It is difficult to distinguish seasonal influenza from other upper respiratory tract infections by symptoms alone, so studies based purely on reported symptoms may not yield a good estimate for the $\R_0$ of seasonal influenza. Instead, studies based on serologically confirmed infections are more reliable.}. One study found $\R_0=1.70$ \citep{Fergu+06}, while another found $\R_0=1.16$ \citep{BrittBecke00} for this outbreak. We average these two values and take $\R_0=1.4$ as a reasonable estimate of the basic reproduction number for seasonal influenza in a modern US population.

\subsection{Vaccination decision model}
Let $p_n$ be the proportion of the population that vaccinates in year $n$. Our goal in this section is to derive a discrete map for the vaccine coverage $p_{n+1}$ in year $n+1$.

In the following argument, we assume the vaccine is fully quantified by its cost and its success rate. For the purposes of this discussion, we will think of the cost in terms of vaccine morbidity (side effects to immunization), though the cost could be interpreted as an economic one (for instance, if individuals have to pay for the vaccine or take unpaid time off of work to obtain it). We denote the cost, or probability of vaccine morbidity, by $0 \leq r \leq 1$, and the probability of vaccine success by $0 \leq s \leq 1$.

To ease the notation in the derivation below, it proves useful to introduce a function $f(\phi)$ to denote the proportion of all \emph{susceptible} individuals who get sick during an epidemic of size $\phi$. To calculate $f$ in terms of $\phi$, note that the fraction of the total population that is susceptible in year $n$ is $1-sp_n$. Of these individuals, a fraction $f\cdot (1-sp_n)$ will get infected, by definition of $f$. But since this fraction also equals the number of infected individuals divided by the total population, it simply equals $\phi$, the fractional size of the epidemic, as given by \eref{phi-expl}. Therefore, $\phi = f \cdot (1-sp_n)$, from which we conclude that
\begin{equation}
   f(\phi(sp_n)) = \frac{\phi(sp_n)}{1-sp_n}.
\end{equation}
In other words, the proportion of susceptible individuals who end up infected is simply the final size of the epidemic renormalized to the susceptible population.

With these preliminaries out of the way, we can deduce the vaccine coverage rate $p_{n+1}$ in year $n+1$ by considering all possible outcomes for an individual based on their choice of whether or not to vaccinate in year $n$, and by counting the proportion of the population flowing down each of the branches in \fref{decision-flowchart} into vaccinating in year $n+1$. We assume that every individual is susceptible to that year's flu strain at the start of each flu season, so everyone must make the choice of whether or not to vaccinate each year.

First consider the group of non-vaccinating individuals, which make up a proportion $1-p_n$ of the population. These individuals will only vaccinate in year $n+1$ if they get sick in year $n$, an event that occurs to a fraction $f = f(\phi(sp_n))$ of them. Thus, the equation for $p_{n+1}$ will include a term $f \cdot (1-p_n)$, which accounts for those that did not vaccinate and got sick.

For vaccinating individuals, either the vaccine succeeds, with probability $s$, or it does not, with probability $1-s$. If the vaccine succeeds, there is still an independent chance that the individual will have side effects that discourage them from vaccinating the following year, which occurs at the vaccine morbidity rate, $r$. However, those for whom the vaccine successfully conferred immunity and provoked no side effects will once again vaccinate the following year since they have no reason to change strategy, which adds the term $(1-r)\cdot s \cdot p_n$ to the equation for $p_{n+1}$.

If the vaccine fails for an individual (with probability $1-s$), but did not cause any discouraging side effects (with probability $1-r$), the only reason they would continue to vaccinate would be if they thought the vaccine succeeded; that is, they happened not to get sick, even though they were not successfully immunized. A proportion $1-f$ of susceptible individuals avoid infection, so the final term of the equation for $p_{n+1}$ is $(1-f)\cdot(1-r)\cdot(1-s) \cdot p_n$.

Putting all of these contributions together, we find that the discrete map for $p_{n+1}$, the proportion of the population vaccinating in year $n+1$, is given by
\begin{equation}
p_{n+1} = \frac{\phi(sp_n)}{1-sp_n}(1-p_n) + (1-r)sp_n + \left[1-\frac{\phi(sp_n)}{1-sp_n}\right](1-r)(1-s)p_n,
\elab{model-map}
\end{equation}
where the function $\phi$ is given by \eref{phi-expl}.
This map is biologically sensible; if $0 \leq p_n \leq 1$, one can check that $0 \leq p_{n+1} \leq 1$. Hence, as long as the initial condition is sensible ($0 \leq p_0 \leq 1$), all subsequent iterations remain in $[0,1]$.

\section{Results}
\slab{results}

\subsection{Model predictions}

The predictions of the model depend on the relative magnitudes of its parameters: the vaccine parameters (morbidity or cost, $r$, and success, $s$), and the disease parameter, $\R_0$. The basic reproduction number $\R_0$ gives a sense of the ``infectiousness'' of the disease; in our analysis, we estimate the basic reproduction number of seasonal influenza in a modern US population to be $\R_0 = 1.4$ (see \sref{math-model} for details), indicating that a person infected with seasonal influenza will infect on average 1.4 other people in a fully susceptible population. For the vaccine parameters, we note that seasonal flu vaccines have very low morbidity \citep{CDCfluvaxsafety}, and their success varies from low to moderate \citep{Oster+12, WHOfluvaxQA, CDC170216}.

The best case would be for the population to self-organize into herd immunity, by driving the proportion of the population vaccinated above the critical threshold, $\pcrit$. When vaccine coverage meets or exceeds this threshold, no epidemic occurs. One might expect the model to self-organize into herd immunity if the population can collectively make use of the memory of the previous flu seasons in a lasting way.

To our disappointment, we find that if there is any cost to the vaccine ($r>0$), our model cannot self-organize into lasting herd immunity. There are two main regions of parameter space in this case \frefp{2param-bifdiag}: a large region where the system eventually converges to vaccine levels below the herd immunity threshold (region I), and a smaller region where the system oscillates in and out of the herd immunity region on a yearly basis (region II).

\begin{figure}
\centering
\includepdfw{0.8}{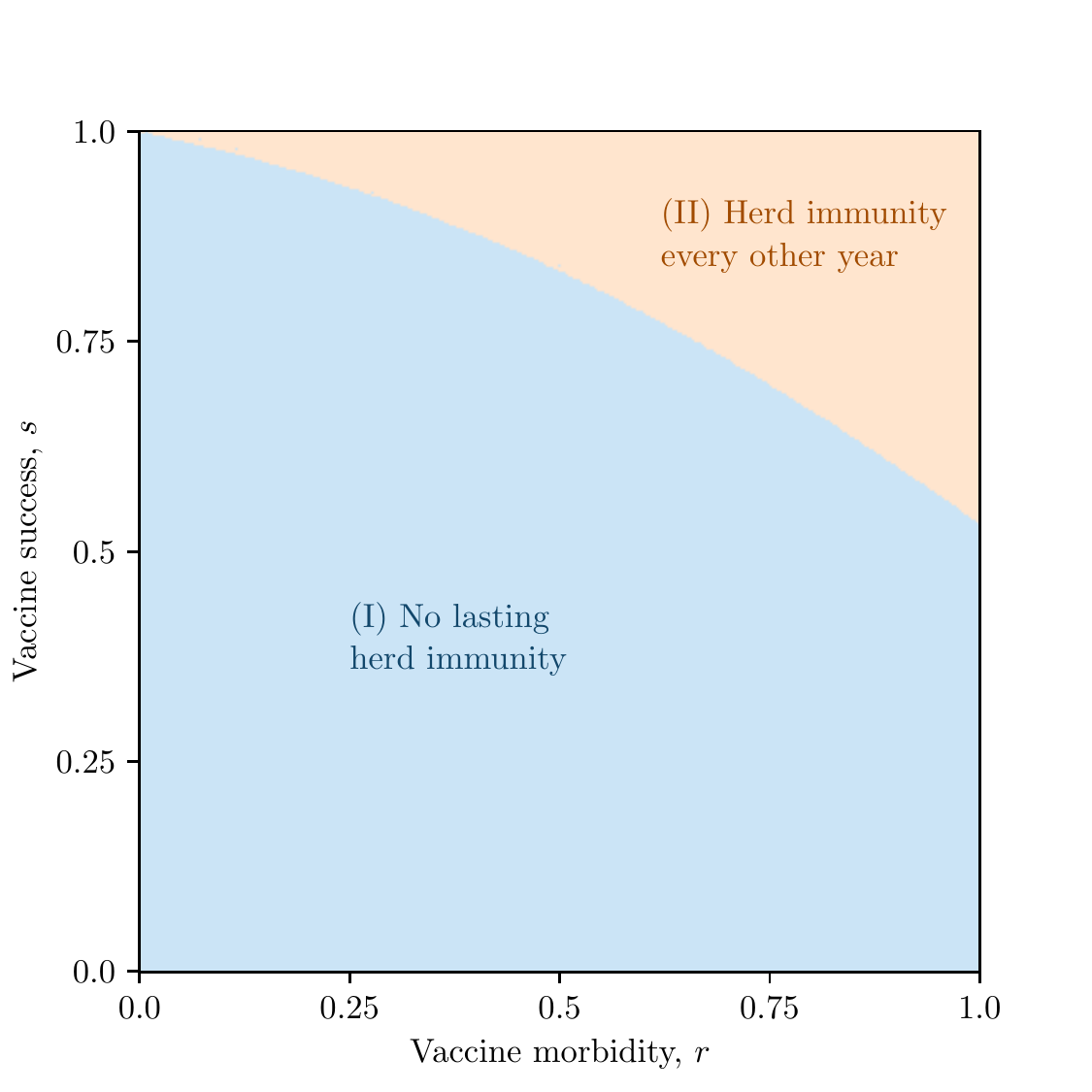}
\caption{\textbf{Long-term model behavior for $\bm{\R_0=1.4}$ and $\bm{p_0 = 0}$, as a function of vaccine morbidity ($\bm{0 < r \leq 1}$) and vaccine success ($\bm{0 \leq s \leq 1}$).} Behavior in these regions of parameter space was deduced by iterating the vaccine coverage map (\eref{model-map}) numerically until it converged to a fixed point. The majority of parameter space is dominated by convergence to vaccine levels below the herd immunity threshold, which results in no lasting herd immunity (region I: the system converges to a period 1 fixed point, $p^*$, that satisfies $p^* < \pcrit$). For higher vaccine success, there is a possibility of achieving herd immunity every other year, provided that vaccine morbidity $r$ is sufficiently large (region II: the system converges to a period-2 fixed point, $(p_1^*, p_2^*)$). In this regime, the system oscillates between sub-optimal vaccine coverage ($p_1^* < \pcrit$) and herd immunity with overvaccination ($p_2^* > \pcrit$): see \fref{oscillating-timeseries}.}
\flab{2param-bifdiag}
\end{figure}

Note that where game-theoretic models always predict free-riding that make herd immunity impossible to achieve in the long term, our model predicts that it is possible for the population to achieve herd immunity every other year, even if there is a cost and moderate failure rate to the vaccine. This oscillatory behavior \frefp{oscillating-timeseries} is the result of the system converging to a state where it alternates between the population bearing a significant disease burden (a large epidemic in the previous year encouraging vaccination in the following year) and the population bearing a significant vaccine cost (which incentivizes non-vaccination \emph{en masse} in the following year). When a vaccine has a moderate-to-high success rate, and a sufficiently high cost, the system is constantly balancing an illness-vaccine cost tradeoff.

\begin{figure}
\floatbox[{\capbeside\thisfloatsetup{capbesideposition={left,center},capbesidewidth=0.25\textwidth}}]{figure}[\FBwidth]
{\caption{\textbf{Vaccine coverage level over time in the regime where herd immunity eventually occurs every other year ($\bm{\R_0=1.4, r=0.55, s=0.9}$).} The system converges to a state where the vaccine coverage level oscillates asymmetrically about the critical vaccination threshold, $p = \pcrit$, denoted by the dashed line.} \flab{oscillating-timeseries}}
{\includepdfw{0.75}{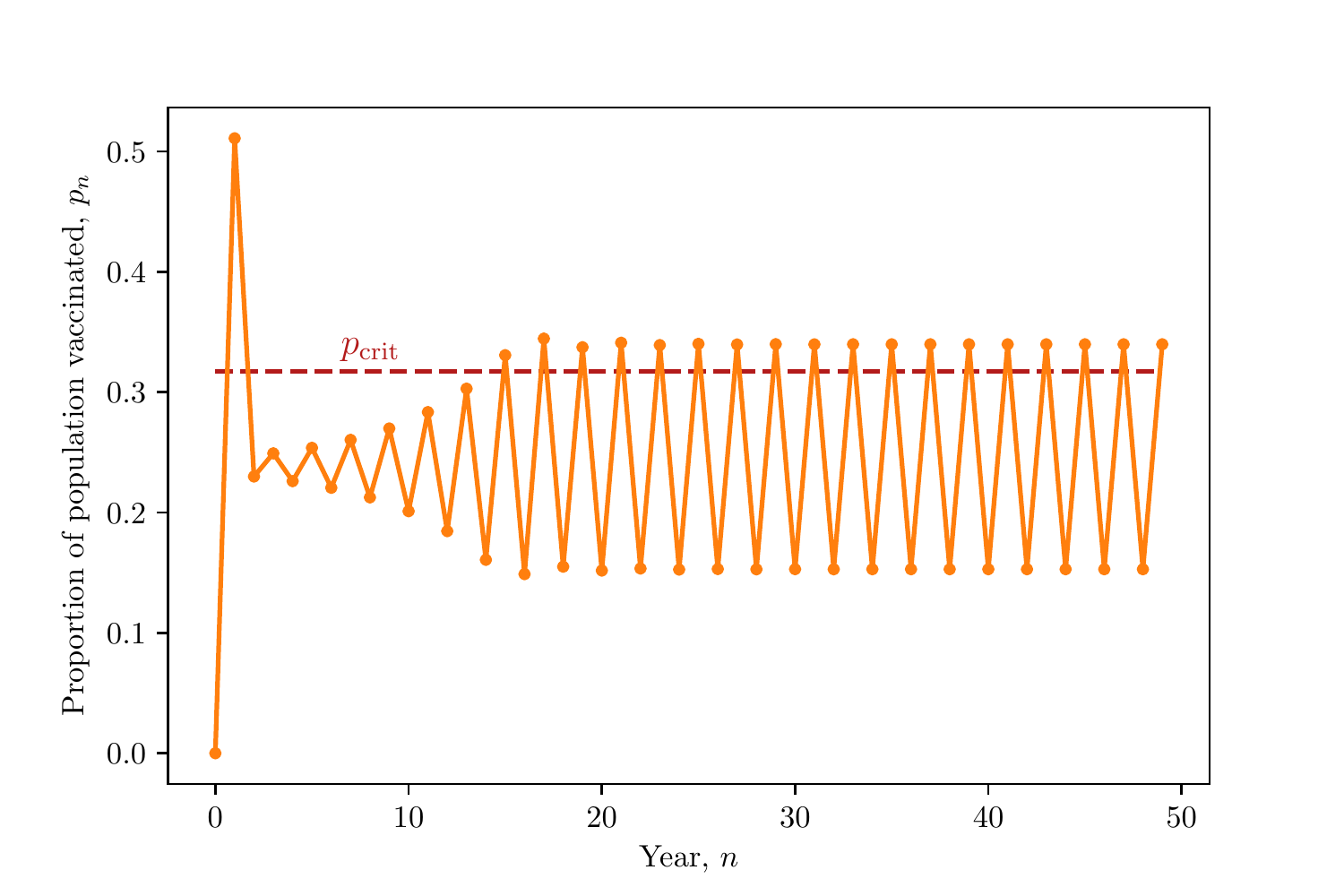}}
\end{figure}

Even in the case where there is no \emph{lasting} herd immunity, the system may nevertheless spend a significant length of time in the herd immunity interval before eventually dropping out \frefp{long-HI-transients}. This effect is especially pronounced when vaccine morbidity is low, and even when the vaccine is only moderately successful, both of which are properties of the real seasonal influenza vaccines. The transient herd immunity period increases as vaccine success increases and/or vaccine cost decreases.

\begin{figure}
\floatbox[{\capbeside\thisfloatsetup{capbesideposition={right,center},capbesidewidth=0.25\textwidth}}]{figure}[\FBwidth]
{\caption{\textbf{Number of years spent in the herd immunity interval ($\bm{p > \pcrit}$) during the transient period in the regime of no lasting herd immunity ($\bm{\R_0 = 1.4}$, $\bm{p_0 = 0}$).} As the vaccine improves in quality (either vaccine morbidity decreases, or vaccine success increases), the time period spent in the herd immunity interval lengthens.}\flab{long-HI-transients}}
{\includepdfw{0.75}{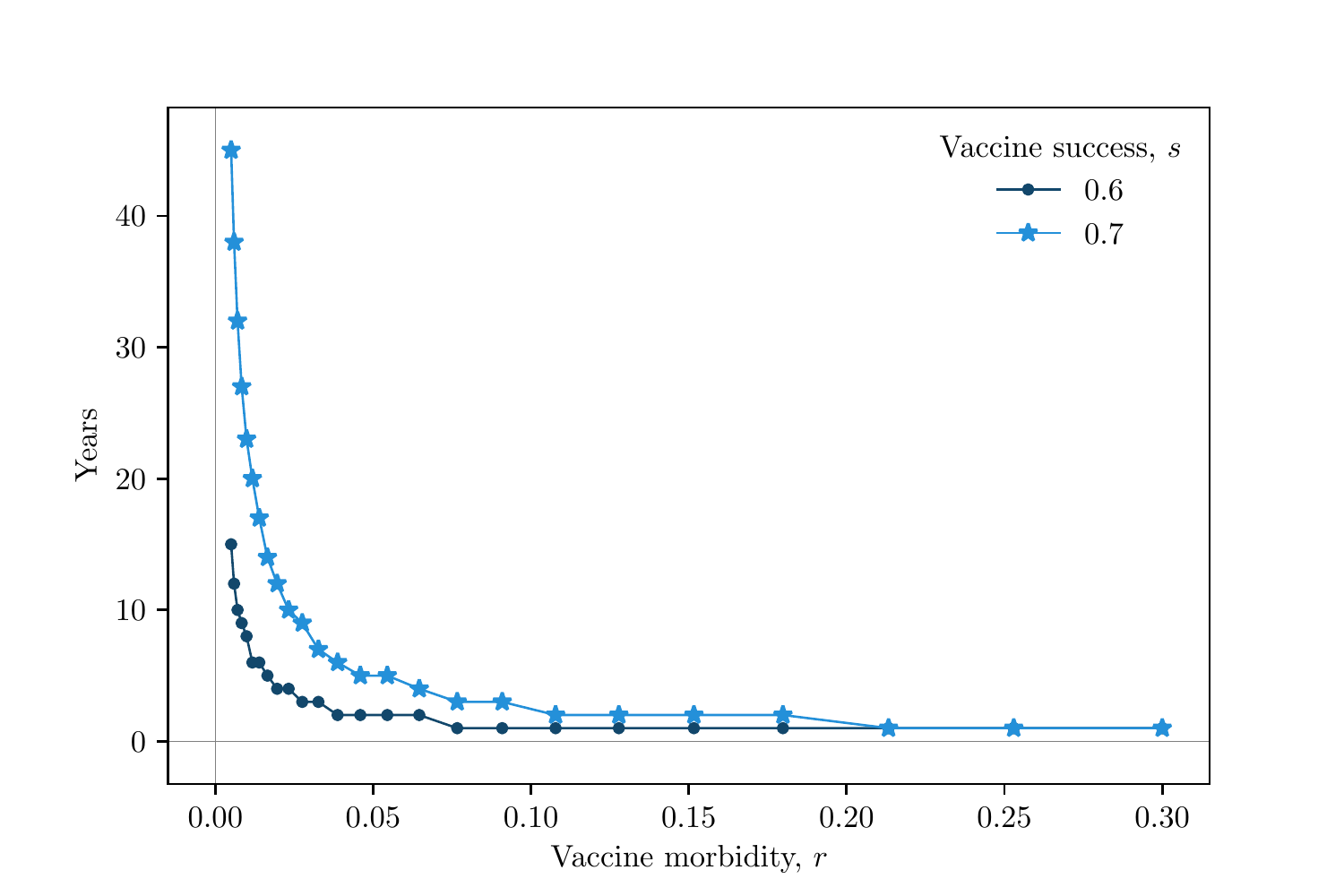}}
\end{figure}

When there is no cost to the vaccine ($r=0$; \fref{SOHI-basins}), the system can self-organize into herd immunity in three ways (regions whose label contains ``lasting herd immunity''), in addition to yielding no lasting herd immunity as before (region I). The system may start in the herd immunity region and therefore stay in it indefinitely (region II), since there is no vaccine cost to drive the coverage level down. Alternatively, the system may converge to lasting herd immunity which is either inefficient as it involves overvaccination (region III), or it may converge to optimal, lasting herd immunity precisely at the herd immunity threshold (region IV).

\begin{figure}
\centering
\includepdfw{0.8}{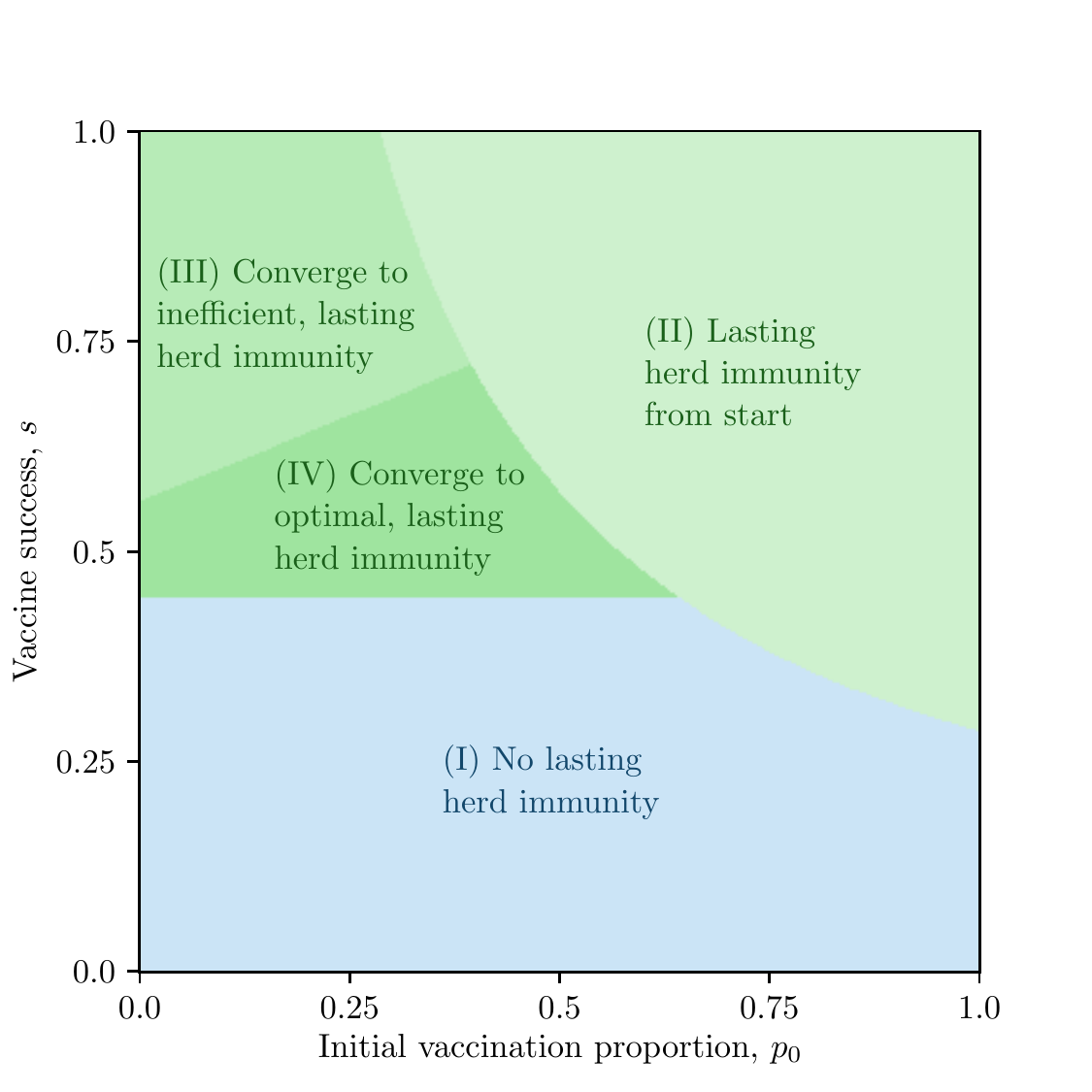}
\caption{\textbf{Long-term model behavior with no vaccine cost ($\bm{r=0}$) for $\bm{\R_0 = 1.4}$.} Behavior in these regions of parameter space can be deduced by iterating the vaccine coverage map (\eref{model-map}) numerically until it converges to a fixed point, but the regions correspond to the analytical criteria detailed in the Appendix. There is still a region with no lasting herd immunity (region I: the system converges to a fixed point, $p^*$, that satisfies $p^* < \pcrit$); in this regime, the system never achieves herd immunity ($p_n < \pcrit$ for all $n \geq 0$). However, the system exhibits self-organized herd immunity when vaccine success is sufficiently high, through a variety of mechanisms. The system may start (and therefore stay) in the herd immunity interval (region II: $p_0 \geq \pcrit$), it may converge to ``inefficient'' lasting herd immunity (region III: sustained overvaccination), or it may converge to ``optimal'' lasting herd immunity (region IV: vaccination approaching the herd immunity threshold $\pcrit$).}
\flab{SOHI-basins}
\end{figure}

The mechanisms that drive the population to either inefficient or optimal self-organized herd immunity are markedly different \frefp{springboard-timeseries}. In the case of sustained overvaccination, the population starts at a relatively low level of vaccination initially (lighter curve). A substantial epidemic occurs in the first year, encouraging a large proportion of individuals to vaccinate in the following year: too many, in fact. The population springs itself into the herd immunity interval after that first year, and since there is no opposing force pushing vaccination coverage down, overvaccination continues indefinitely. In the case of optimal vaccination, the population may also start at a relatively low level of initial vaccination, but the first epidemic sustained is not as devastating as in the previous case (darker curve). A moderate proportion of the population is affected by the disease and switches to vaccinating in the following year. The epidemic sustained in this next year is not as large as the one before it (thanks to the increase in vaccination), and encourages another (smaller) group of individuals to switch to vaccinating next year. This process gradually guides the population to the herd immunity threshold, eventually achieving the optimal level of vaccination.

\begin{figure}
\floatbox[{\capbeside\thisfloatsetup{capbesideposition={right,center},capbesidewidth=0.25\textwidth}}]{figure}[\FBwidth]
{\caption{\textbf{Vaccine coverage level over time with no vaccine cost ($\bm{r=0}$) in the regime of self-organized herd immunity $\bm{(\R_0=1.4, s=0.6})$.} If the initial population level is too low (lighter curve), the population springs into the interior of the herd immunity interval, $[\pcrit, 1]$, which results in sustained overvaccination. If the population initially vaccinates at a more moderate level (darker curve), vaccine coverage converges to the optimal herd immunity threshold, $\pcrit$, in an asymptotic way.}\flab{springboard-timeseries}}
{\includepdfw{0.75}{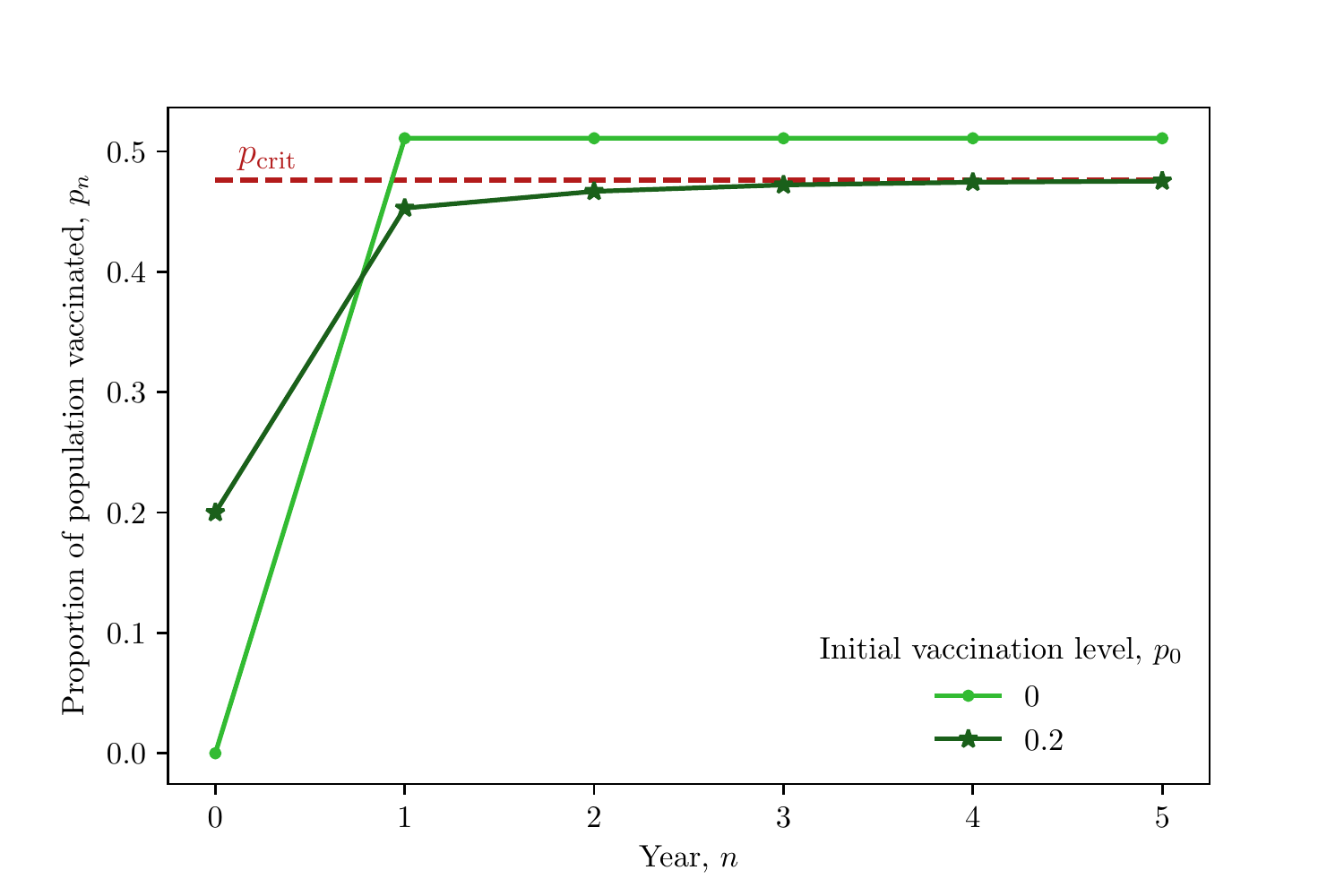}}
\end{figure}

\section{Discussion}
\slab{discussion}

While the simple model studied here cannot self-organize into sustained herd immunity when there is any cost to the vaccine, it may still achieve herd immunity every other year. When the vaccine success rate is sufficiently large for a given cost, there is an ongoing battle between the disease and the vaccine. If the population undervaccinates in one year, it undergoes an epidemic which drives the system to overvaccinate in the following year. A non-trivial proportion of the population then bears some cost associated with the vaccine, which discourages those individuals from getting vaccinated the following year, driving the system back down to an undervaccinated state, and the cycle repeats. Provided the amplitude of this (asymmetric) oscillation about the herd immunity threshold is sufficiently small, this regime effectively achieves herd immunity as any epidemic that occurs is relatively small. While the goal of disease eradication has not strictly been achieved, the resulting epidemics are so small in the model that a bit of stochasticity may be enough to push the circulating flu strain into extinction (in a closed population).

Although this promising biannual behavior is possible, the region of no lasting herd immunity dominates vaccine parameter space, particularly for vaccine morbidity and success levels that are realistic for the seasonal influenza vaccine (cost near zero, success rate around 50\%) \frefp{2param-bifdiag}. The system may not drive itself to herd immunity asymptotically for these types of vaccines, but a significant length of time is spent in the herd immunity interval during the transient period \frefp{long-HI-transients}. This effect opens the door for other public health interventions (\eg vaccination- and disease-awareness campaigns) which have not been included in the model but may help push the population into lasting herd immunity. Increases to the length of time spent in the herd immunity interval can be achieved by improving the vaccine, by increasing vaccine success, and/or by decreasing vaccine morbidity.

In the case where there is no cost to the vaccine, the system can achieve self-organized herd immunity that is either inefficient (due to overvaccination) or optimal (at the critical vaccination threshold). If the population initially vaccinates at a very low level, it undergoes a large epidemic, and the following year, the population overreacts, propelling itself into the herd immunity interval much like a diver on a springboard. Overvaccination continues since there is no cost to the vaccine, and thus no force pushing population vaccine coverage down. On the other hand, moderate initial vaccination leads the population to converge to the optimal vaccination level at the herd immunity threshold. In this case, the population gradually learns from year to year through successively smaller epidemics. Each such epidemic recruits smaller and smaller proportions of the population to vaccinate until herd immunity is achieved. Notably, this result occurs even with a moderately effective vaccine, like that of real seasonal influenza.

While it may not be realistic to assume that a vaccine can be considered costless to an individual, this extreme case illustrates that if a vaccine can be perceived as costless, the population can self-organize into sustained herd immunity. Such a result is still possible even if the vaccine is only moderately successful and even if the population does not immediately take to getting vaccinated.

\section{Conclusion}
\slab{conclusion}
We have presented an intentionally simple model for seasonal influenza vaccination that challenges the usual assumption that individuals make use of perfect, global information completely rationally when making annual flu vaccination decisions. The usual assumption wrongly predicts widespread use of free-rider logic, which is not typically observed in this context. We make use of established results in social psychology to inform our model, which gives rise to both interesting and interpretable dynamics.

In the case where there is some cost to the vaccine, our model still predicts regimes where vaccination coverage is below the herd immunity threshold in the long term, which agrees with previous models. However, our model also predicts new regions where herd immunity is achieved every other year: a result of the population oscillating between vaccine-based and disease-based morbidity. When we further assume that the vaccine has no cost, it is still possible for the model to predict no lasting herd immunity. We also observe convergence to enduring herd immunity, either at the optimal level, where the population vaccinates exactly enough to reach this protected state, or inefficiently, where the population overvaccinates.

Our disease-behavior model is deliberately simple as a first step, to focus on the effect of incorporating a more realistic decision model on top of a well-established model for disease spread. Future work should focus on making this model more realistic and validating it with appropriate data.

Currently, both the decision-making and disease processes are deterministic; a stochastic version of this model in either respect would be closer to reality. Since agents only rely on the current state of the system to inform their next decision, our model could easily be cast in a Markov chain framework. Modeling the disease spread on a socio-spatial network would also provide greater realism, by mimicking the way hosts interact and thus spread infectious diseases like the flu \citep{Chao+10}.

Seasonal influenza is an immensely complex phenomenon, and we have not accounted for issues such as that of multiple concurrently circulating strains \citep{Prosp+11}, cross-reactivity of vaccines between strains \citep{Iorio+12, Moa+16}, or waning vaccine immunity over the course of the flu season \citep{Rambh+18}. Such additions to the model would also serve to make it more realistic.

\bibliographystyle{spmpscinat} 
\bibliography{Papst_etal_2021_flu-vaccine-decisions}   

\newpage

\section*{Appendix: Analytical criteria for long-term behavior in model with no cost ($r=0$)}
\slab{appendix}

\fref{SOHI-basins} can be produced by directly iterating the map to numerical convergence, but it can be produced equivalently using analytical criteria based on the existence and location of fixed points.

\def\theequation{A\arabic{equation}}
\setlength{\parindent}{0em}
\setlength{\parskip}{1em}

When $r=0$, the model map (\eref{model-map}) reduces to

\begin{equation}
p_{n+1} = \frac{\phi(sp_n)}{1-sp_n}(1-p_n) + sp_n + \left[1-\frac{\phi(sp_n)}{1-sp_n}\right](1-s)p_n.
\end{equation}

Fixed points, $p$, of this map must satisfy
\begin{equation}
p = \frac{\phi(sp)}{1-sp}(1-p) + sp + \left[1-\frac{\phi(sp)}{1-sp}\right](1-s)p,
\end{equation}
which simplifies to
\begin{equation}
\frac{\phi(sp)}{1-sp}(p(2-s)-1) = 0.
\elab{fps-nocost}
\end{equation}
\eref{fps-nocost} is satisfied when either (i) $\phi(sp) = 0$ or (ii) $p(2-s)-1 = 0$.

Case (i) is satisfied by any $p \in [\pcrit, 1]$ since $\phi(sp) = 0$ if and only if $p \geq \pcrit$. Provided $\pcrit < 1$, the "herd immunity" interval $[\pcrit, 1]$ is an invariant set of neutrally stable fixed points that exists in the map's domain of $[0,1]$. In other words, if $p_n \in [\pcrit, 1]$ for any $n$, the trajectory is then trapped in the herd immunity interval for all remaining time (precisely at the value $p_n$).

Region II in \fref{SOHI-basins} is given by all $(p_0,s)$ that satisfy $p_0 \geq \pcrit = \frac{1}{s}\left(1 - \frac{1}{\R_0}\right)$. In other words, the population starts at herd immunity and remains at herd immunity indefinitely. Region III is given by all $(p_0, s)$ such that $p_0 < \pcrit$ but $p_1 \geq \pcrit$. This region represents populations whose first epidemic was so large that it propels the population into herd immunity immediately after the first year.

Case (ii) is satisfied by $p^* = 1/(2-s)$. This fixed point is disjoint from the herd immunity interval when
\begin{align}
    p^* &< \pcrit, \\
    \frac{1}{2-s} &< \frac{1}{s}\left(1-\frac{1}{\R_0}\right), \\
    s &< 1 - \frac{1}{2\R_0 - 1}.
\end{align}
Let us define $\scrit = 1-1/(2\R_0 - 1)$; when $s \geq \scrit$,  the fixed point $p^* = 1/(2-s)$ disappears into the herd immunity interval.

Numerical simulations suggest that $p^*$ is stable when it exists disjoint from the herd immunity interval (\ie when $s < \scrit$), with the basin of attraction being $[0, \pcrit)$; such trajectories will converge to $p^*$ as $n \to \infty$. Region I is given by all $(p_0, s)$ that satisfy $p_0 < \pcrit$ and $s < \scrit$. In other words, vaccine coverage starts below the herd immunity threshold and never surpasses it. Instead, vaccine coverage converges to $p^* = 1/(2-s) < \pcrit$.

Lastly, region IV is given by all $(p_0, s)$ with both $p_0 < \pcrit$ and $p_1 < \pcrit$, but also $s\geq \scrit$. In this case, the fixed point $p = 1/(2-s)$ does not exist distinct from the herd immunity interval, but also the first epidemic is not strong enough propel vaccine coverage over the herd immunity threshold ($p_1 <\pcrit$). Under these conditions, numerical simulations suggest that $\{p_0, p_1, ... \}$ is a monotonically increasing sequence where each $p_n < \pcrit$, but $\lim_{n \to \infty} p_n = \pcrit$.

\end{document}